\documentclass[aps,preprint]{revtex4-1}
\usepackage{graphicx}
\usepackage{amsmath}
\usepackage{makeidx}
\usepackage{amsfonts}
\usepackage{amssymb}

\begin{document}

\title{Computing loop corrections by message passing}

\author{A. Ramezanpour}
\affiliation{Department of Applied Science and Technology, Politecnico di Torino,
Corso Duca degli Abruzzi 24, 10129 Torino, Italy}

\email{abolfazl.ramezanpour@polito.it}

\date{\today}

\begin{abstract}
Any spanning tree in a loopy interaction graph can be used for 
communicating the effect of the loopy interactions by introducing messages that are passed along the edges in the spanning tree. This
defines an exact mapping of the problem on the loopy interaction graph onto an extended problem on a tree interaction graph, where the thermodynamic quantities can be computed by a message-passing algorithm based on the Bethe equations. We propose an approximation loop correction algorithm for the Ising model relying on the above representation of the problem. The algorithm deals at the same time with the short and long loops, and can be used to obtain upper and lower bounds for the free energy.
\end{abstract}


\maketitle

\section{Introduction}
Interesting problems are usually computationally hard and it is always useful to have efficient and accurate approximation algorithms. Computing local probability marginals for an arbitrary Gibbs measure is one of these fundamental problems in statistical physics and computer science. For example, having an accurate estimation of the probability marginals is enough to solve a constrained satisfaction problem by a decimation algorithm.           

There are some exactly solvable models that can be served as starting points for studying the nontrivial corrections in more interesting and complicated problems. The main examples are the mean-field (MF) solution of an infinite dimensional system \cite{BW-prs-1934}, and the Bethe solution of interacting systems with a tree structure \cite{B-prs-1935}. In the former case, we are concerned with the finite dimensional corrections and in the latter, which is the subject of this study, we are concerned with the loop corrections. 

Given a tree interaction graph, the local probability marginals can be computed by the belief propagation (BP) algorithm, which minimizes the Bethe free energy by passing messages (or cavity marginals) along the graph edges \cite{KFL-inform-2001,MP-epjb-2001,MM-book-2009}. The loopy belief propagation algorithm is an extension of the BP algorithm to loopy interaction graphs to find a local minimum of the Bethe free energy, which could be larger or smaller than the exact one; see Refs. \cite{Y-2002,H-2006,WYM-2012} for some attempts to construct a convex free energy approximation and ensure the algorithm convergence. Nevertheless, the main strategy to deal with the loops is to group the variables in larger regions to eliminate some loops of a given length scale as in the cluster variational method \cite{K-pr-1951,P-jphysa-2005} or generalized BP \cite{YFW-nips-2001}. In the extreme limit we have the junction tree method \cite{LS-jrs-1988}, where the regions are chosen large enough to get a tree interaction graph. There are other algorithms that try to approximate a probability measure with simpler ones having tree structures \cite{MQ-2004,WJW-ieee-2003,K-ieee-2006}.   

The local marginals provided by the loopy BP algorithm are not necessarily consistent beyond the two-point correlations, and one way to improve the algorithm is to increase the range of consistent marginals \cite{MR-jstat-2006,MWKR-proc-2007}. On the other hand, any fixed point of the loopy BP equations can be used to obtain a loop expansion of finite but exponential number of terms, starting from the loopy BP contribution \cite{CC-jstat-2006}. In particular, this allows to show that for a class of attractive (or ferromagnetic) models the loopy BP algorithm provides an upper bound for the free energy \cite{SWW-pnip-2007}. Finally, there are some efforts to construct field theories expanded around the Bethe solution \cite{E-physica-1990,PS-jstat-2006}.   

\begin{figure}
\includegraphics[width=6cm,height=7cm]{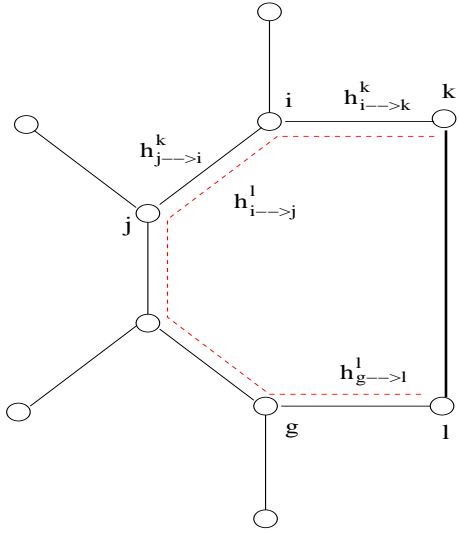}
\caption{Propagating the effect of loopy interactions through a spanning tree.
Here, the loopy interaction $J_{kl}s_ks_l$ is replaced with the two effective fields $h_{i\to k}^k=1/2 J_{kl}s_l$ and $h_{g\to l}^l=1/2 J_{kl}s_k$, which are locally determined by the messages that are passed along the edges in the spanning tree.}\label{f1}
\end{figure}

In this paper we present an approximation loop correction algorithm that is based on the following observation:   
In loopy graphs, a global quantity can be computed locally by decomposing the computation into smaller ones distributed among different elements in the graph and collected by messages that are passed along the edges of any spanning tree \cite{Mackay-book,RRZ-epjb-2011,RZ-prb-2012}. Consider the Ising model on an arbitrary graph and one of its spanning trees. Given a spin configuration, one can provide to each spin the effective field originated from the loopy interactions by passing some messages through the spanning tree. The messages are updated at each node to collect the effective fields coming from different parts of the tree; see Fig. \ref{f1}. In this way, we obtain an exact mapping of the spin configurations on the loopy graph onto the larger space of the spin and effective field configurations on the spanning tree. Of course, this mapping does not change the problem complexity but it offers some approximation loop correction algorithms relying on the above representation.  

In the following we discuss more about the details and write the loop correction equations for the Ising model. Then, we study some approximations to reduce the algorithm complexity by considering only a relevant subset of the loopy interactions and treating the other ones in a mean-field approximation. This provides an upper bound for the free energy. We also obtain a lower bound for the free energy by a convex combination of the loopy interactions. 

\section{Loop corrections in the Ising model}
Consider the Ising model with Hamiltonian $H=-\sum_{i=1}^N B_i s_i-\sum_{(ij)\in E} J_{ij}s_is_j$ on the interaction graph $G=(V,E)$ with the set of nodes $V$ and edges $E$. Given a spanning tree $\mathsf{T}=(V,E_0)$ and spin configuration $\underline{s} \in \{-1,+1\}^N$, we rewrite the Hamiltonian as 
\begin{align}
H=-\sum_{i=1}^N (B_i+B_i^{\mathsf{T}}) s_i-\sum_{(ij)\in E_0} J_{ij}s_is_j,
\end{align}  
with $B_i^{\mathsf{T}} \equiv  \frac{1}{2}\sum_{j \in \partial i \setminus \partial_0 i } J_{ij}s_j$. 
Here $\partial i$ and $\partial_0 i$ denote the neighborhood set of $i$ in $G$ and $\mathsf{T}$, respectively. The set $E \setminus E_0$ defines the set of loopy interactions with respect to the spanning tree $\mathsf{T}$. In words, $B_i^{\mathsf{T}}$ is the effective field resulted from the loopy interactions. We are going to write the local fields $B_i^{\mathsf{T}}$ in terms of some messages that are  propagated through the spanning tree. To account for a loopy interaction $(kl)\in E \setminus E_0$  
we need to pass the messages $(h_{i\to j}^{l}=\frac{1}{2}J_{kl}s_k,h_{j\to i}^{k}=\frac{1}{2}J_{kl}s_l)$ on edges $(ij)$ in the unique path $k \leftrightarrow l = (k, \dots  ,i, j, \dots,l)$ connecting $k$ and $l$ on $\mathsf{T}$. Figure \ref{f1} shows how a single loopy interaction $J_{kl}s_ks_l$ can be represented by the effective fields $(h_{i\to k}^{k},h_{g\to l}^{l})$, which are locally determined by the messages along the path $k \leftrightarrow l$. Collecting all the messages that are resulted from the loopy interactions we obtain:
\begin{align}
h_{i\to j}^l= \sum_{k \in \partial_0 i \setminus j} h_{k \to i}^l+\frac{1}{2} J_{il}s_i(1-\delta_{l,j}) \equiv \hat{h}_{i\to j}^l.
\end{align}  
Consequently, we can write $B_i^{\mathsf{T}}= \sum_{j \in \partial_0 i} h_{j \to i}^i$.
Note that on each directed edge $(i\to j)\in \mathsf{T}$ we have a vector of messages $\vec{h}_{i \to j}=\{h_{i \to j}^l| l \in \mathsf{T}_{j\to i} \}$. The cavity tree $\mathsf{T}_{i\to j}$ is defined recursively by $i \cup \{\mathsf{T}_{k\to i}|k \in \partial_0 i \setminus j\}$. 

Finally, considering the fact that the messages $\vec{h}_{i \to j}$ are uniquely determined by the spin configuration, we write the Ising partition function as
\begin{align}
Z=\sum_{\underline{s}} e^{-H}=\sum_{\underline{s}}  \int \prod_{i=1,\dots,N}\mathbb{I}_{h}^{(i)}\prod_{j\in \partial_0 i}  d\vec{h}_{i \to j} 
e^{\sum_{i=1}^N (B_i+\sum_{j \in \partial_0 i} h_{j \to i}^i) s_i+\sum_{(ij)\in E_0} J_{ij}s_is_j}.
\end{align}  
The indicator function $\mathbb{I}_{h}^{(i)}\equiv \prod_{j \in \partial_0 i} \prod_{l\in \mathsf{T}_{j\to i}} \delta(h_{i\to j}^l-\hat{h}_{i\to j}^{l})$ ensures that the sum over the messages $h_{i\to j}^l$ is one when the messages satisfy the equations $\hat{h}_{i\to j}^{l}$, otherwise it is zero. 

Now, the interaction graph is a tree and we can compute the free energy and the local marginals by the Bethe equations. These are recursive equations for the cavity marginals $\mu_{i \to j}(s_i;\vec{h}_{ij})$, that is the probability of having spin $s_i$ and messages $\vec{h}_{ij} \equiv (\vec{h}_{i\to j},\vec{h}_{j\to i})$ in absence of node $j$. Notice that we need the variables $(s_i,\vec{h}_{ij})$ to determine recursively the equilibrium properties of the spins in the cavity tree $\mathsf{T}_{i\to j}$. One can write $\mu_{i \to j}(s_i;\vec{h}_{ij})$ in terms of the neighboring cavity marginals $\{\mu_{k \to i}(s_k;\vec{h}_{ik})|k\in \partial_0 i \setminus j\}$, considering also the effect of the local soft and hard interactions. The equations governing the cavity marginals, called belief propagation (BP) equations \cite{KFL-inform-2001}, read
\begin{multline}\label{BPh}
\mu_{i \to j}(s_i;\vec{h}_{ij}) \propto   \int \prod_{k\in \partial_0 i \setminus j} d\vec{h}_{ik} \times \mathbb{I}_{h}^{(i)}
e^{(B_i+\sum_{k\in \partial_0 i}h_{k \to i}^i) s_i} \prod_{k\in \partial_0 i \setminus j} \left(\sum_{s_k}e^{ J_{ik}s_is_k} \mu_{k \to i}(s_k;\vec{h}_{ik}) \right).
\end{multline}
There is, of course, one and only one solution to the BP equations that can be found by iteration starting from the leaves. In this case we do not need to specify the initial conditions as the above equations define the cavity marginals one after the other. Or, one can start from random initial messages $\mu_{i \to j}(s_i;\vec{h}_{ij})$ and update them in a random sequential way according to the BP equations; the tree structure of the interaction graph ensures that the algorithm converges to the unique fixed point of the equations.          
Then the free energy is computed by the Bethe form of the free energy: $F=\sum_{i} \Delta F_i-\sum_{(ij) \in E_0} \Delta F_{ij}$, where $\Delta F_i$ and $\Delta F_{ij}$ are the local free energy changes by adding node $i$ and link $(ij)$ to the interaction graph \cite{MM-book-2009}, i.e.,
\begin{align}
e^{-\Delta F_i} &=\sum_{s_i} \int \prod_{k\in \partial_0 i} d\vec{h}_{ik} \times
\mathbb{I}_{h}^{(i)} e^{(B_i+\sum_{k\in \partial_0 i}h_{k \to i}^i) s_i} 
\prod_{k\in \partial_0 i} \left(\sum_{s_k}e^{ J_{ik}s_is_k} \mu_{k \to i}(s_k;\vec{h}_{ik}) \right), \\
e^{-\Delta F_{ij}} &=\sum_{s_i,s_j}\int d\vec{h}_{ij}  
e^{ J_{ij}s_is_j} \mu_{i \to j}(s_i;\vec{h}_{ij})\mu_{j \to i}(s_j;\vec{h}_{ij}).
\end{align}  
Similarly, one can compute the local marginals $\mu_i(s_i)$ and $\mu_{ij}(s_i,s_j)$. Moreover, by $\langle 2h_{j\to i}^is_i \rangle$ one obtains the average energy of the loopy interactions connecting node $i$ to the nodes in the cavity tree $\mathsf{T}_{j\to i}$.   

\begin{figure}
\includegraphics[width=8cm]{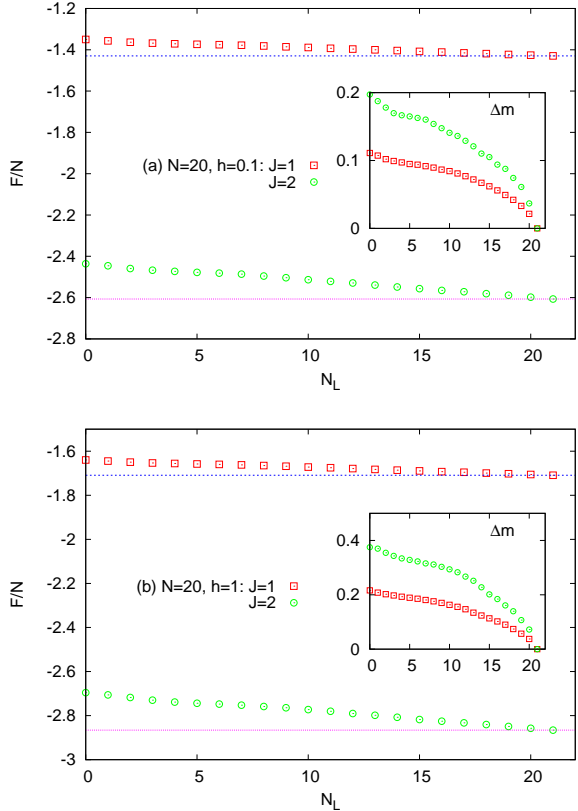}
\caption{The exact free energy $F$ and magnetization difference $\Delta m\equiv 1/N \sum_i|m_i^{L}-m_i^{exact}|$ vs the number of the loopy interactions added to the maximum spanning tree in a $2D$ square lattice of size $N=4\times 5$. We take random Gaussian fields $B_i$ of mean zero and variance $h^2$, and random Gaussian couplings $J_{ij}$ of mean zero and variance $J^2$. The points are averaged over $500$ independent realizations of the random fields and couplings for $h=0.1$ (a), and  $h=1$ (b). The errorbars are smaller than the size of points.}\label{f2}
\end{figure}
  
\section{Considering a subset of the loopy interactions}
The time and memory complexity of the above algorithm (in the worst case) grow exponentially with the number of loopy interactions and we have to resort to some reasonable approximations. Here, we focus on a class of approximations that work with a subset of the loopy interactions and preserve the upper bound property of the free energy. The approximation performance then depends on the structure of the spanning tree and the subset of the loopy interactions.

In practice, one can start with a maximum weight $W\equiv \sum_{(ij)\in \mathsf{T}} |J_{ij}|$ spanning tree and add the loopy interactions one by one according to some criterion. Then at each step one obtains an upper bound for the free energy of the original interacting system after adding the energy contribution of the discarded loopy interactions; see Fig. \ref{f2}. The problem complexity would depend on the number of nonzero vector elements in the $\vec{h}_{i \to j}$. Let us define $X_{i\to j}$ as the set of spins in the cavity tree $\mathsf{T}_{i\to j}$ that interact by loopy interactions with some spins in the cavity tree $\mathsf{T}_{j\to i}$. Then, the vector $\vec{h}_{i \to j}$ takes $2^{|X_{i\to j}|}$ values corresponding to the effective fields for different spin configurations in $\mathsf{T}_{i\to j}$ that are relevant for the spins in $\mathsf{T}_{j\to i}$. 

Let us start with a mean-field approximation of the loopy interactions, where we take the following effective Hamiltonian
$H_{eff}=-\sum_{i=1}^N (B_i+\langle B_i^{\mathsf{T}} \rangle) s_i-\sum_{(ij)\in E_0} J_{ij}s_is_j$. In words, we replace the effective fields of the loopy
interactions with their average values. Equivalently, it means that we update the messages according to $h_{i\to j}^l= \sum_{k \in \partial_0 i \setminus j} h_{k \to i}^l+\frac{1}{2} J_{il}m_i(1-\delta_{l,j})$. Given the average magnetizations $m_i$, the BP equations for the effective system are written in terms of the cavity marginals $\mu_{i \to j}(s_i)$,     
\begin{align}
\mu_{i \to j}(s_i) \propto e^{(B_i+\frac{1}{2}\sum_{k\in \partial i \setminus \partial_0 i} J_{ik}m_k) s_i} 
\prod_{k\in \partial_0 i \setminus j} \left(\sum_{s_k}e^{ J_{ik}s_is_k} \mu_{k \to i}(s_k) \right).
\end{align}
This is the probability of observing state $s_i$ for spin $i$ in the cavity tree $\mathsf{T}_{i\to j}$, that is neglecting the interaction with spin $j$.  
The magnetizations $m_i$ are determined self-consistently by the local marginals
\begin{align}
m_i=\frac{1}{Z_i} \sum_{s_i} s_i e^{(B_i+\frac{1}{2}\sum_{k\in \partial i \setminus \partial_0 i} J_{ik}m_k) s_i} 
\prod_{k\in \partial_0 i} \left(\sum_{s_k}e^{ J_{ik}s_is_k} \mu_{k \to i}(s_k) \right),
\end{align} 
where $Z_i$ is the normalization constant. The cavity marginals and the magnetizations can be found by iteration starting from random initial values and updating them in a random sequential way. Note that the algorithm may not converge after adding the self-consistent equations for the magnetizations. Still, if the algorithm converges one can use the cavity marginals to obtain an upper bound for the free energy. In fact, the Gibbs measure induced by the  effective Hamiltonian $H_{eff}$ can be considered as a variational probability distribution for the system. Since the interaction graph is a tree we are sure that we compute the effective free energy and entropy exactly. Therefore, by adding the average energy of the discarded loopy interactions and subtracting the average energy of the effective interactions $\sum_i \langle B_i^{\mathsf{T}} \rangle s_i$ we obtain an upper bound for the free energy of the original system.     
  
Now, suppose we have partitioned the set of the loopy interactions into two subsets $\mathcal{L}_{E}$ and $\mathcal{L}_{MF}$ that are to be treated exactly and in the mean-field approximation, respectively. The subset $\mathcal{L}_{E}$ could be the collection of all the loopy interactions within the $r$th neighborhood of the nodes in $\mathsf{T}$ which we denote by $\mathcal{L}_r$. For $r=0$ we recover the above mean-field approximation. Then, we rewrite the BP equations as
\begin{multline}\label{BPh-r}
\mu_{i \to j}(s_i;\vec{h}_{ij}(r)) \propto   \int \prod_{k\in \partial_0 i \setminus j} d\vec{h}_{ik}(r) \times \mathbb{I}_{h}^{(i)}
e^{\left(B_i+\frac{1}{2}\sum_{(ik)\in \mathcal{L}_{MF}(i) } J_{ik}m_k+\sum_{k \in \partial_0 i}h_{k \to i}^i(r) \right) s_i} \\ \times \prod_{k\in \partial_0 i \setminus j} \left(\sum_{s_k}e^{ J_{ik}s_is_k} \mu_{k \to i}(s_k;\vec{h}_{ik}(r)) \right),
\end{multline} 
where $\vec{h}_{i\to j}(r)$ is the vector of messages going to the nodes $k \in \mathsf{T}_{j\to i}$ with distance $d_{ik}\le r$. And $\mathcal{L}_{MF}(i)$ is the set of the loopy interactions involving $i$ that are treated in the mean-field approximation.  Similarly, one can write the self-consistent equations for the magnetizations. The equations can be solved by iteration starting from random initial cavity marginals and magnetizations. There is, of course, no guarantee that the algorithm converges after introducing the MF approximation. 
But, as explained above, if the algorithm converges the cavity marginals can be used to obtain an upper bound for the free energy.

\begin{figure}
\includegraphics[width=8cm]{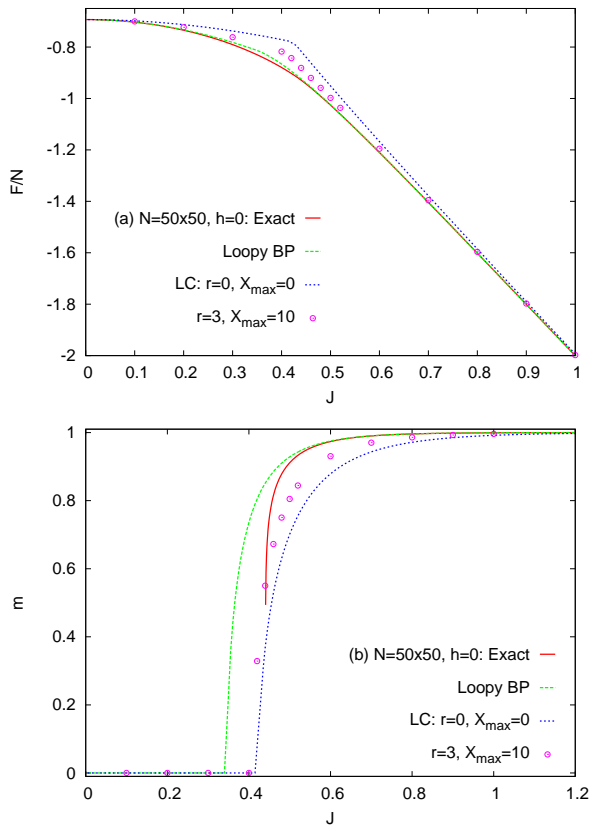}
\caption{(a) The free energy $F$, and (b) the magnetization $m$ vs the strength of the couplings in a $2$D square lattice of size $N=50\times 50$ at zero external fields $B_i=0$ with ferromagnetic couplings $J_{ij}=J$. The loopy BP algorithm (dashed line) and the exact thermodynamic solution (solid line) are compared with the loop correction (LC) algorithm. The LC results have been obtained with a random spanning tree for different values of $r$, the length scale of the short loops, and $X_{max}$, the computational complexity of the algorithm determined by the structure of the loopy interaction graph.}\label{f3}
\end{figure}

Note that for an arbitrary subset $\mathcal{L}_{E}$, the computation time scales as $N2^{Y_{max}}(K_{max}X_{max})^2$ with $K_{max}$ the maximum degree in the spanning tree, $X_{max}$ is the maximum $|X_{i\to j}|$, and $Y_{max}\equiv \max_{i} |\cup_{j \in \partial_0 i}X_{j\to i}|$. We can indeed control the algorithm complexity by adding the loopy interactions as long as $X_{max}$ is smaller than a given value.

\begin{figure}
\includegraphics[width=8cm]{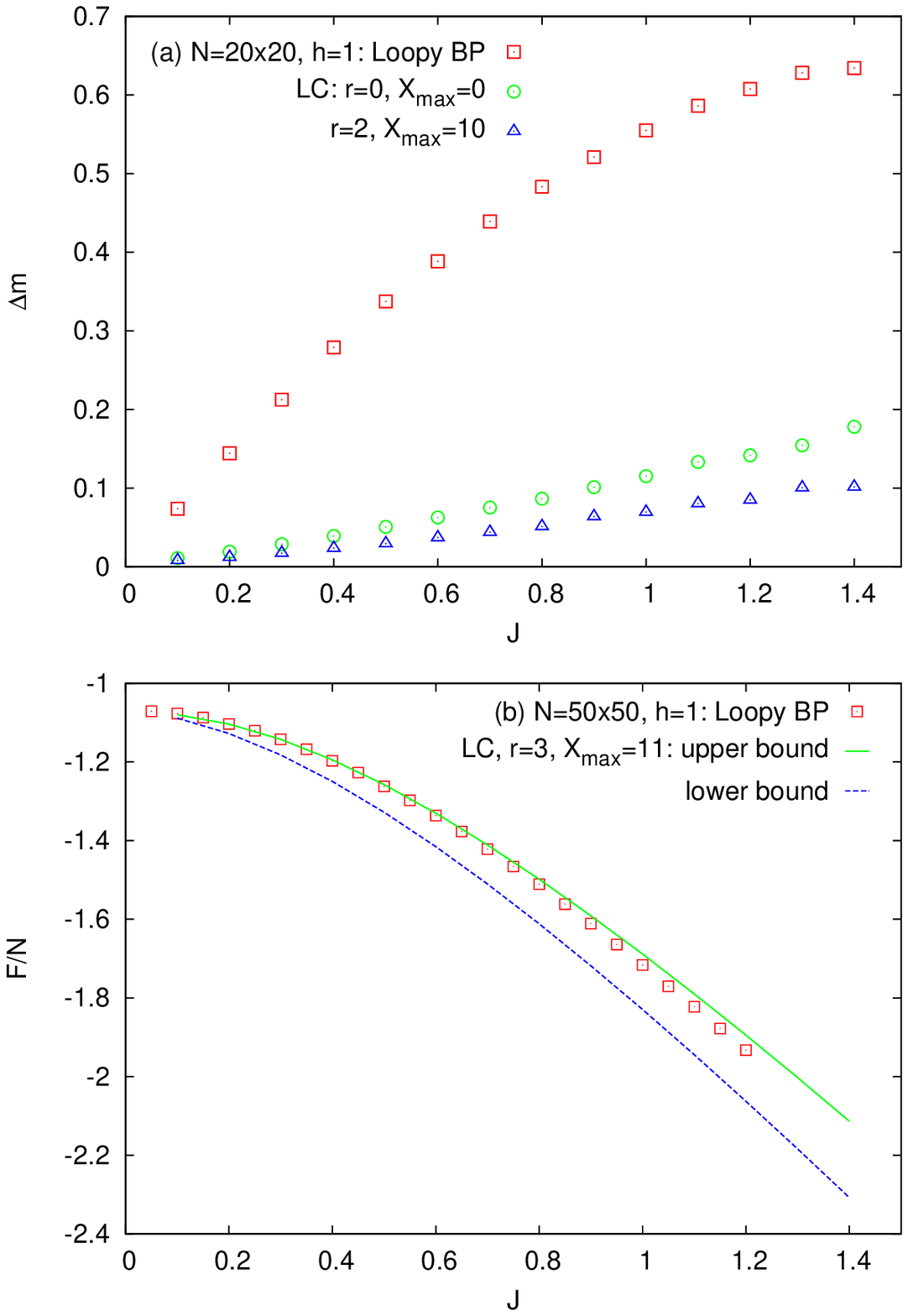}
\caption{(a) The magnetization difference $\Delta m\equiv 1/N \sum_i |m_i^{LC}-m_i^{MC}|$ (relative to the Monte Carlo results), and (b) the free energy upper and lower bounds vs the strength of the couplings in a $2$D square lattice with random Gaussian fields $B_i$ of mean zero and variance $h^2$, and random Gaussian couplings $J_{ij}$ of mean zero and variance $J^2$. The loopy BP algorithm is compared with the loop correction (LC) algorithm based on the maximum spanning tree, for different values of $r$ and $X_{max}$. The data points in (a) are averaged over at least $10$ independent realizations of the random fields and couplings. The lower and upper bounds in (b) have been obtained for a single instance of the problem whereas the loopy BP results are averaged over at least $10$ independent realizations. The errorbars are smaller than the size of points.}\label{f4}
\end{figure}

In Figs. \ref{f3} and \ref{f4} we compare the algorithm performance with the loopy BP algorithm in a two-dimensional ($2$D) square lattice. For the subset $\mathcal{L}_{E}$ we have chosen the first $N_L$ loopy interactions of largest magnitudes plus the short loopy interactions in the local neighborhood of radius $r$, such that $|X_{i\to j}| \le X_{max}$ for all the directed links. Notice that the free energy obtained by the loopy BP is very good as long as the algorithm converges, however, in general we do not know if this free energy is smaller or larger than the exact one. Usually it is more difficult to provide this information and at the same time give a good estimation of the free energy. Moreover, as illustrated in the figures, the loopy BP algorithm is not doing well in predicting the local magnetizations in the disordered system with random fields and couplings, and close to the transition point in the homogeneous system.

\section{A convex combination of the loopy interactions}
One can use the above algorithm to obtain some lower bounds for the free energy. The free energy is a concave function of the fields and the couplings, therefore, $F[\mathbf{B},\mathbf{J}]\ge \sum_{g} P_g F[\mathbf{B}^g,\mathbf{J}^g]$ for any probability measure $P_g$ over the parameters $(\mathbf{B}^g,\mathbf{J}^g)$ as long as $\mathbf{B}=\sum_{g} P_g\mathbf{B}^g$ and $\mathbf{J}=\sum_{g} P_g\mathbf{J}^g$. The couplings $\mathbf{J}^g$ are chosen such that the free energy $F[\mathbf{B}^g,\mathbf{J}^g]$ can be computed exactly. Let us assume that the interaction graphs $G_g$ defined by $\mathbf{J}^g$ are loopy graphs of complexity less than $X_{max}$ with respect to a spanning tree $\mathsf{T}$. Finding the optimal lower bound is difficult but any consistent set of the interactions and the measure $P_g$ give a lower bound for the free energy. For instance, the tree-reweighted algorithm works with the spanning trees $G_g$ and uses message passing techniques to address the above optimization problem \cite{WJW-ieee-2003,K-ieee-2006}. 

Here, we take the maximum weight spanning tree $\mathsf{T}$, and partition the set of the loopy interactions $\mathcal{L}$ into subsets $\mathcal{L}_r$ and $\{\mathcal{L}_g|g=1,\dots,\mathcal{N}\}$. We recall that $\mathcal{L}_r$ contains all the loopy interactions within the $r$th neighborhood of the nodes in $\mathsf{T}$. Then we add the loopy interactions in $\mathcal{L}_r$ and $\mathcal{L}_g$ to the spanning tree to obtain  $G_g=\mathsf{T} \cup \mathcal{L}_r \cup \mathcal{L}_g$. Moreover, for the sake of simplicity, we assume the fields are the same in all the sub-problems $B_i^g=B_i$, and $J_{ij}^g=J_{ij}$ when the link $(ij)$ belongs to the subset $\mathsf{T} \cup \mathcal{L}_r$. The other loopy interactions for $(ij) \in \mathcal{L}_g$ are set to $J_{ij}^g=J_{ij}/P_g$.  A reasonable choice for the probability of having interaction graph $G_g$ is $P_g= W_g/(\sum_{g'} W_{g'})$, where $W_g\equiv \sum_{(ij)\in \mathcal{L}_g} |J_{ij}|$. Then as long as the sub-problems have a small complexity $X_{max}$ we can use our loop correction algorithm to compute exactly the free energies $F[\mathbf{B}^g,\mathbf{J}^g]$ and so a lower bound for the free energy $F[\mathbf{B},\mathbf{J}]$. Figure \ref{f4} displays the lower bound that we obtain in this way for the free energy in a $2$D square lattice with random fields and couplings.   

\section{Conclusion}
In summary, we introduced an approximation message-passing algorithm to compute loop corrections in the Ising model. The approximation works with a relevant subset of the loopy interactions and uses the mean-field approximation to deal with the other loopy interactions. Obviously, the algorithm in this form is more suited to systems with strongly heterogeneous couplings. It would be very useful to have other approximation algorithms which treat all the loopy interactions in the same manner. And finally, the algorithm can be used to obtain better lower bounds for the free energy than the naive one that we presented here.

\acknowledgments
I would like to thank A. Bruanstein, M. Mezard, A. Pagnani, and R. Zecchina for helpful discussions. Support from ERC Grant No. OPTINF  267915 is acknowledged.

\end{document}